\documentclass{article}
\usepackage[utf8]{inputenc}
\usepackage{textcomp}
\DeclareUnicodeCharacter{207B}{\textsuperscript{-}}

\PassOptionsToPackage{numbers, compress}{natbib}
\usepackage[dblblindworkshop, final]{neurips_2026}

\workshoptitle{I Can't Believe It's Not Better (ICBINB): Failure Modes of AI in Biology}

\usepackage{float}
\usepackage[utf8]{inputenc}
\usepackage[T1]{fontenc}
\usepackage{hyperref}
\usepackage{url}
\usepackage{booktabs}
\usepackage{amsfonts}
\usepackage{amsmath}
\usepackage{amssymb}
\usepackage{microtype}
\usepackage{enumitem}
\setlist{nosep,leftmargin=*}
\usepackage{xcolor}

\usepackage{graphicx}
\usepackage{multirow}
\usepackage{textgreek}
\usepackage{natbib}

\title{Protein Language Models Fail To Capture Contact Predictions Under Invariant Relabeling}

\author{%
  Paarth Batra\thanks{Equal contribution.} \\
  The Biodesign Institute \\
  Tempe, AZ 85281, USA \\
  \texttt{pbatra6@asu.edu} \\
  \And
  Aryan Keluskar\footnotemark[1] \\
  The Biodesign Institute \\
  Tempe, AZ 85281, USA \\
  \texttt{akeluska@asu.edu} \\
}

\begin{document}

\maketitle

\begin{abstract}
Protein language models (pLMs) produce highly accurate contact maps from single-sequence inputs. People often take this accuracy as proof that these models implicitly learn invariant coevolutionary statistics, extracting the same signals as classical alignment-based methods like Potts. 
In this work, we introduce a diagnostic framework around a global bijective relabeling of the amino-acid alphabet. This transformation permutes residue letters while leaving all alignment-level information, column entropies, and pairwise edit distances unchanged. Pseudolikelihood Potts models and inverse-covariance estimators remain invariant under this transformation, but pLMs fail the test. Across multiple architectures (ESM-2, MSA Transformer, AMPLIFY) and model scales (8M to 3B parameters), pLM contact maps collapse under arbitrary relabeling.
Using Hamming distance controls and biochemical grouping interventions, we attempt to isolate what likely causes this degradation and find that these models rely on explicit, canonical residue identities (``alphabet anchoring'') rather than pure coevolutionary statistics. Increasing scale makes models more responsive to biochemical classes, but their performance still degrades under relabeling. 
Our findings show a gap between how pLMs and classical coevolutionary models compute contacts. More broadly, our work shows how diagnostic perturbations that leave classical models provably unchanged can expose hidden representation biases and prevent over-reliance on neural black boxes in biological modeling.
\end{abstract}

\section{Introduction}

\begin{figure}[t]
    \centering
    \includegraphics[width=\textwidth]{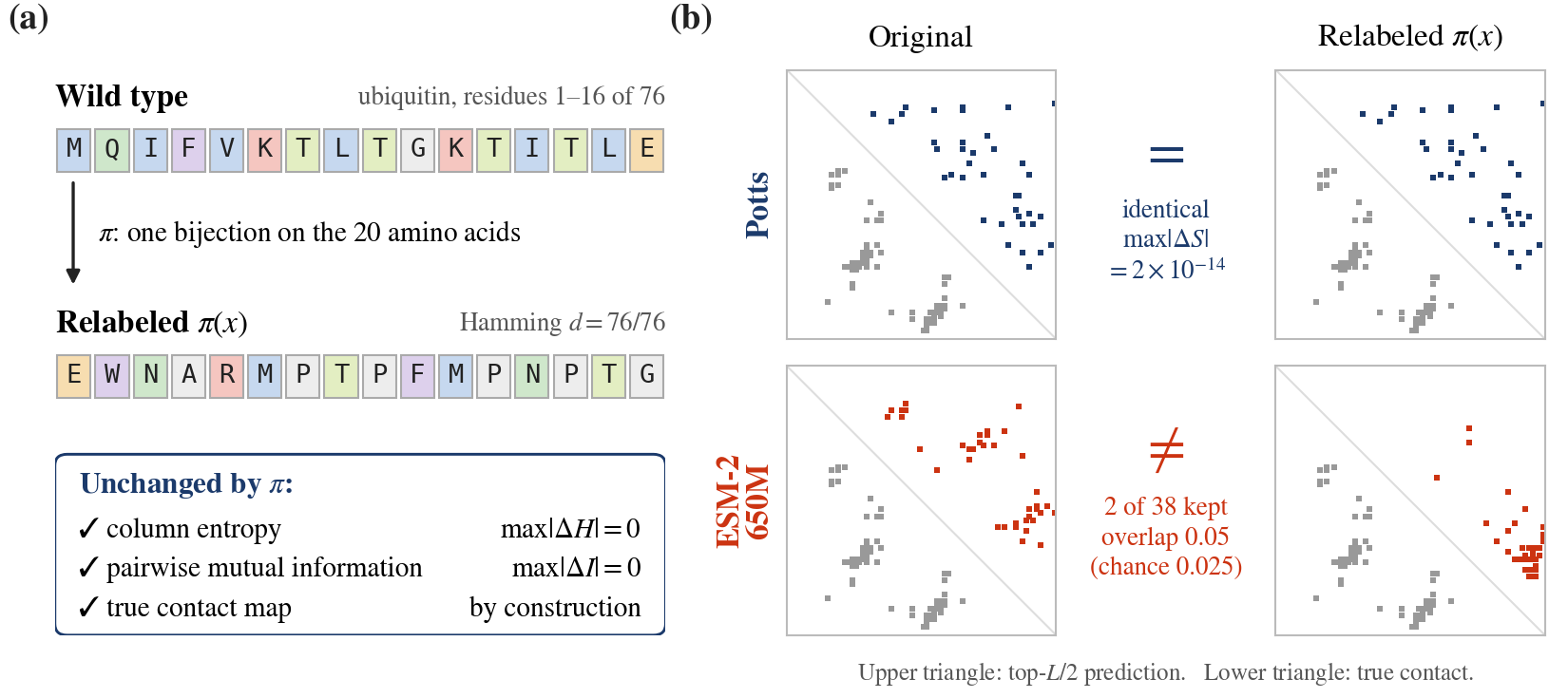}
    \caption{\textbf{A global relabeling of the amino-acid alphabet leaves every alignment statistic intact and destroys the pLM's contact map.} \textbf{(a)} The intervention: one bijection $\pi$ on the twenty amino acids, applied to every residue of every sequence. Both $\Delta$s are measured on $4{,}000$ sequences of the ubiquitin family alignment. \textbf{(b)} The result on the same input. Lower triangle: long-range contacts of the deposited chain (1UBQ\_A); upper triangle: each model's own top-$L/2$ prediction, over the candidate pairs the alignment covers. Potts returns a bit-identical score matrix (max$|\Delta S| = 2\times10^{-14}$, Proposition~2), while ESM-2 650M keeps $2$ of its own $38$ contacts (overlap $0.05$, chance $0.025$).}
    \label{fig:condition}
\end{figure}

Protein language models (pLMs) trained on masked-residue prediction alone recover structural and functional information from a single sequence, with no alignment in the pipeline \citep{rives2021biological}. ESM-2 extends this to atomic-resolution structure prediction at speeds orders of magnitude beyond alignment-based methods \citep{lin2023evolutionary}. Contact maps are the standard, canonical way to show that pLMs encode structure \cite{vig2020bertology}. Their accuracy is widely interpreted as evidence that the models have internalized coevolution, and therefore learned the signal that direct-coupling analysis extracts explicitly from a multiple sequence alignment (MSA) by fitting a Potts model \citep{morcos2011direct}. Further work shows that ESM-2 stores coevolutionary statistics in a coupling tensor \citep{zhang2024protein}.

However, two mechanisms can produce high contact precision on natural proteins. A model may learn to invert the covariance of the alignment's column statistics, the frequencies of residues at each position and of residue pairs across positions, or it may pattern-match on which amino-acid letters occupy each position. On a leakage-clean split, contact readouts lose 30--36 precision points, which shows that letter identity carries predictive signal independent of covariance \citep{thorstenson2026protein}. Since precision alone cannot distinguish these mechanisms, the right diagnostic \citep{vig2020causal} is a test that the hypothesized mechanism must pass regardless of the model's accuracy \citep{geirhos2020shortcut}.

In this work, we test the coevolution interpretation by renaming every amino acid consistently across all sequences in a family. The renaming moves each column's statistics onto new letters without changing their values, so any estimator that reads only column statistics must return the same contacts. We prove this for the pseudolikelihood Potts model \citep{ekeberg2013improved} and, in closed form, for the inverse-covariance estimator \citep{jones2012psicov}. By the contrapositive, a model whose contact predictions change under relabeling is not computing them from column statistics alone. Unlike shuffle controls \citep{rao2021msa} or reduced alphabets \citep{ieremie2024protein}, a bijective relabeling loses no information \citep{cover2006elements}. To separate edit distance from biochemistry, we add relabelings confined to groups of similar residues \citep{dayhoff1978model, henikoff1992amino} and a matched-Hamming control that substitutes randomly at exactly the positions a relabeling changes.

The classical Potts estimator \citep{marks2011protein} passes this test, and every pLM we examine fails(Table~\ref{tab:main}, Figure~\ref{fig:condition}). Averaged over nine proteins, ESM-2 650M keeps an overlap of only $0.015$ with its original contacts once all twenty letters are relabeled at random; MSA Transformer and AMPLIFY fail the same way. This failure is not caused by input difficulty: a relabeling confined to similar groups keeps an overlap of $0.655$. On cytochrome~c, random substitutions at relabeled positions destroy the map. Larger models tolerate biochemically similar swaps better, but arbitrary relabeling is as destructive across the board(Figure~\ref{fig:cost}a). The failure also costs accuracy: on ubiquitin, over a candidate set where the two precisions are directly comparable, ESM-2 goes from $2.33\times$ the Potts precision to $0.17\times$ under relabeling. Contact precision therefore cannot be cited as proof that a pLM computes contacts through coevolution.

To summarize, our contributions in this work are as follows:
\begin{itemize}
    \item We introduce a global relabeling of the amino-acid alphabet as a test of whether a contact readout uses only column statistics, and prove that pseudolikelihood Potts models and inverse-covariance estimators pass it (\S\ref{sec:condition}).
    \item We show that every pLM we test fails it, at every ESM-2 size, and that for ESM-2 650M the failure replicates on its supervised contact head and on variant-effect prediction (\S\ref{sec:rq1}, \S\ref{sec:rq3}).
    \item Using biochemistry-preserving relabelings and a matched-Hamming control, we separate the contribution of residue identity from edit distance, and term the phenomenon of models' reliance on residue identity "\emph{alphabet anchoring}" (\S\ref{sec:rq2}, \S\ref{sec:discussion}).
\end{itemize}

\section{Background and Related Work}

\textbf{Coevolutionary contact prediction.}
Classical contact prediction reads structure from the second-order column statistics of a multiple sequence alignment (MSA). By fitting a maximum-entropy Potts model \citep{weigt2009identification, morcos2011direct} or inverting a regularized covariance matrix \citep{jones2012psicov}, these methods separate direct coevolution between two positions from indirect correlation through others. Protein language models (pLMs) compute no alignment. Their contacts can be read from the categorical Jacobian $J[i, a, j, b]$, the sensitivity of the prediction at position $j$ for amino acid $b$ to setting position $i$ to amino acid $a$ \citep{thorstenson2026protein, zhang2024protein}. The Jacobian has the same shape as the Potts coupling tensor, and the same post-processing turns either into a contact map: mean-centering, symmetrization, the Frobenius norm over amino-acid pairs, and average product correction (APC) \citep{dunn2008mutual, marks2011protein}. This shared form is the basis for the claim that pLMs implicitly learn coevolution. Appendix~\ref{app:preliminaries} gives the full pipeline.

\textbf{Evaluation beyond accuracy.}
Prior evaluations of pLM representations often rely on downstream accuracy benchmarks \citep{dieckhaus2025protein, ding2024protein, yang2024convolutions}, which cannot tell a model that computes the intended quantity from one that relies on the distribution of natural sequences. Following out-of-distribution diagnostics in general language modeling \citep{zhao2026chain}, we shift the input in a way that changes no information in the alignment's column statistics. Because the classical baseline is exactly invariant to this shift, the test does not depend on the choice of benchmark (\S\ref{sec:ood}).

\textbf{Input controls for contact prediction.}
\citet{rao2021msa} probed MSA-based models along two axes: shuffling values within a column, which destroys covariation, and shuffling the order of columns, which destroys position. They also hypothesized that deeper transformer layers become invariant to amino-acid identity. Reduced amino-acid alphabets \citep{ieremie2024protein} change column entropy and discard information, so a failure under reduction shows only that the input got harder. However, none of these controls changes residue identity while leaving the alignment's statistics intact. We add that axis: a bijective relabeling of the alphabet, which loses no information \citep{cover2006elements}, and which tests at the level of contacts the invariance that \citet{rao2021msa} hypothesized.

\section{Methodology}

We investigate whether the contact readout of a protein language model behaves like a function of the alignment's column statistics. Our methodology centers on three research questions:

\begin{itemize}

    \item \textbf{RQ1.} Does a protein language model's contact prediction remain invariant under a global bijective relabeling of the amino-acid alphabet?

    \item \textbf{RQ2.} When a neural model's contact prediction degrades under relabeling, is the degradation driven by the disruption of biochemical similarity classes or by the increase in edit distance?

    \item \textbf{RQ3.} Does this degradation shrink with model scale, and does it cost accuracy against solved structures?

\end{itemize}

\subsection{Classical Estimators Are Invariant to a Global Relabeling}
\label{sec:condition}

A global bijection on the twenty standard amino acids moves each column's statistics onto new letters without changing their values. Since column entropies, inter-column mutual information, and the true contact map are all unchanged by construction, any function of column statistics must be exactly invariant under the bijection. We prove this for two estimators in Propositions~1 and~2, with full proofs in Appendix~\ref{app:proofs}.

The inverse-covariance estimator satisfies the condition exactly (Proposition~2), and it remains accurate on real proteins such as ubiquitin. A pseudolikelihood Potts fit under L-BFGS returns an identical top-$L/2$ contact set from an equivariant initialization \citep{seemayer2014ccmpred} and $0.970$ overlap from a random one (Appendix~\ref{app:c13}). Both conditions of the theorem are needed: making the ridge anisotropic breaks the invariance (max$|\Delta S| = 0.355$), and so does letting the gap symbol enter the permuted set ($0.182$), against about $10^{-15}$ when the conditions hold.

\subsection{Intervention Design}
\label{sec:regimes}

A relabeled sequence differs from the original at many positions, so a degradation could come from edit distance alone. We therefore use three regimes, each moving $m$ of the twenty residue types, to separate edit distance from biochemical class.

\emph{Random} is a uniform derangement of $m$ residue types. \emph{In-class} is confined to BLOSUM62-similar groups (IVLM, KR, DE, NQ, ST, FYW), preserving biochemistry while changing identity \citep{murphy2000simplified}. \emph{Cross-class} moves each residue it changes into a different group, at the same edit budget. Figure~\ref{fig:regimes} shows the permutation maps. At nominal $m=20$, in-class moves only $15$ residue types, because A, C, G, H and P have no BLOSUM62 partner, and cross-class moves $14$ to $16$ (Appendix~\ref{app:details}). Tables and text give nominal $m$; figures plot each point at its realized $m$.

We also add a \emph{matched-Hamming} control. For each (protein, regime, $m$, seed) we build the permuted sequence alongside a uniform random substitution and a composition-matched substitution at the same set of positions, so all three arms mutate the same number of residues and differ only in which letters they install (Appendix~\ref{app:c6}).

\subsection{Models, data, and metrics}\label{sec:setup}

We test ESM-2 at five sizes from 8M to 3B parameters \citep{lin2023evolutionary}, MSA Transformer \citep{rao2021msa}, and AMPLIFY 120M~\citep{fournier2024protein}, against an inverse-covariance Potts baseline and a pseudolikelihood Potts fit. MSA Transformer reads an alignment at inference exactly as Potts does, and neither is retrained, removing the asymmetry that Potts is refit on the permuted alignment while ESM-2 receives no relabeled training data. AMPLIFY has a different vocabulary ($27$ tokens in a different residue order), a different feed-forward design (SwiGLU), and a different training corpus, at matched capacity. All experiments ran on a single NVIDIA A100-80GB GPU.

We report two metrics. \emph{Overlap} is the fraction of a model's top-$L/2$ long-range contacts on the original sequence that it also predicts on the relabeled one, so an invariant readout scores $1$. \emph{Precision} P@$L/2$ is measured against solved structures \citep{monastyrskyy2014evaluation}. We fixed each experiment's protocol before running it, including the outcome that would refute it, and we report the tests that did not meet their thresholds as such (\S\ref{sec:ood}, \S\ref{sec:negative}). Protein lists, structures, and per-cell counts are in Appendix~\ref{app:details}.

\section{Results}

\subsection{RQ1: Are Protein Language Model Contact Maps Invariant to Relabeling?}
\label{sec:rq1}

No protein language model we test is invariant (Table~\ref{tab:main}). Averaged over nine proteins and three seeds, ESM-2 650M's overlap under random relabeling falls to $0.015$ at $m=20$, while in-class relabeling keeps $0.655$ at its realized $m=15$. The cross-class arm falls faster than the random arm at $m=6$ and $m=10$, which \S\ref{sec:discussion} explains.

\begin{table}[ht]
  \centering
  \small
  \begin{tabular}{llcccc}
    \toprule
    Model & Regime & $m=2$ & $m=6$ & $m=10$ & $m=20$ \\
    \midrule
    \multirow{3}{*}{ESM-2 650M} & random & $0.702$ & $0.354$ & $0.100$ & $\mathbf{0.015}$ \\
    & in-class & $0.879$ & $0.809$ & $0.684$ & $\mathbf{0.655}$ \\
    & cross-class & $0.570$ & $0.247$ & $0.027$ & $0.026^\dagger$ \\
    \midrule
    \multirow{2}{*}{MSA Transformer} & random & $0.775$ & $0.471$ & $0.244$ & $\mathbf{0.147}$ \\
    & in-class & $0.935$ & $0.795$ & $0.640$ & $\mathbf{0.536}$ \\
    \midrule
    \multirow{2}{*}{AMPLIFY 120M} & random & $0.559$ & $0.222$ & $0.124$ & $\mathbf{0.056}$ \\
    & in-class & $0.770$ & $0.671$ & $0.493$ & $\mathbf{0.414}$ \\
    \midrule
    Potts (inverse covariance) & any & \multicolumn{4}{c}{$1.000$} \\
    \bottomrule
  \end{tabular}
  \vspace{10pt}
  \caption{Contact-set overlap under alphabet relabeling; $1$ means the model predicts the same contacts. Columns give nominal $m$: at $m=20$, in-class moves $15$ residue types and cross-class $14$ to $16$. ESM-2 650M: nine proteins, three seeds, $n=27$ per cell ($\dagger$: $n=9$). MSA Transformer: four Pfam families, three seeds ($n=12$). AMPLIFY 120M: three proteins, two seeds ($n=6$). Potts is bit-identical by Proposition~2.}
\label{tab:main}
\end{table}

MSA Transformer reads the same relabeled alignments that Potts is fit to. On those alignments Potts returns identical contacts in all $136$ cells (Appendix~\ref{app:details}), while MSA Transformer falls to $0.147$ under random relabeling and keeps $0.536$ in-class. AMPLIFY reproduces both the collapse and the in-class--random split, and the in-class to random ratio widens with $m$. Unlike ESM-2 and MSA Transformer, which share a $33$-token vocabulary and training corpus, AMPLIFY has its own $27$-token vocabulary in a different residue order, its own training corpus, and a different feed-forward design (SwiGLU) at matched capacity. Under random relabeling at $m=20$ it loses every contact in its own top-$L/2$ on cytochrome~c. The degradation is therefore not specific to one model family or tokenizer. Figure~\ref{fig:contacts} shows the contact maps.

\subsection{RQ2: Does Biochemical Class or Edit Distance Drive the Degradation?}
\label{sec:rq2}

\begin{figure}[t]
  \centering
  \includegraphics[width=\textwidth]{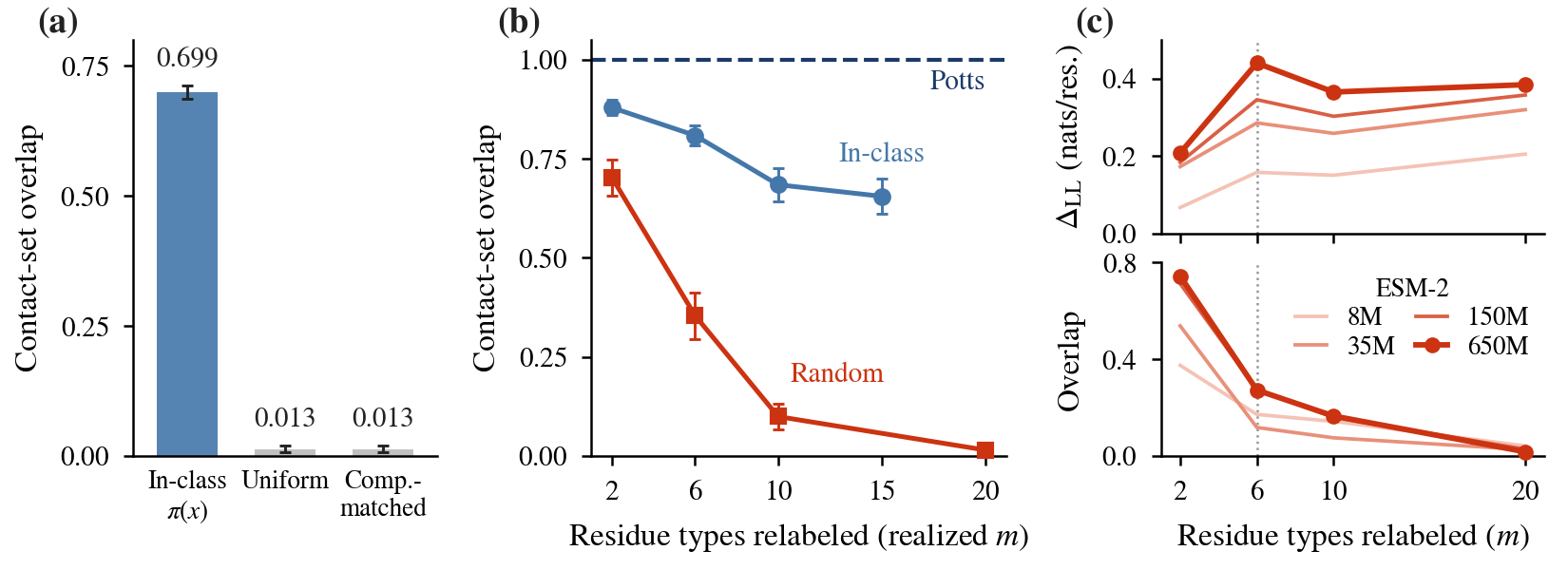}
  \caption{\textbf{Overlap tracks biochemical class at matched Hamming distance, and keeps falling after likelihood shift saturates.} \textbf{(a)} On cytochrome~c the three arms change the same $76$ of $104$ positions and differ only in which residues they install. In-class overlap is $0.699$; both matched substitutions give $0.013$. \textbf{(b)} Potts is the flat line at $1.0$ by Proposition~2. ESM-2 650M separates by biochemical class, and the cross-class arm falls faster than the random arm. \textbf{(c)} In the random arm, $\Delta_{\mathrm{LL}}$ saturates by $m{=}6$ while overlap falls another $16\times$. Faint curves are 8M, 35M, and 150M; bold curves are 650M.}
  \label{fig:mechanism}
\end{figure}

Figure~\ref{fig:mechanism}(a) holds edit distance fixed. On cytochrome~c, in-class relabeling at nominal $m=20$ changes $76$ of $104$ positions for all three seeds and keeps overlap $0.699$. Uniform random and composition-matched substitutions at the same $76$ positions both give $0.013$, so the relabeled sequence keeps $54\times$ more of its own contact set than either control. At matched edit distance, the outcome depends on which letters are installed: substitutions within biochemical groups cost little, and arbitrary substitutions destroy the prediction.

Figure~\ref{fig:mechanism}(b) shows the same separation across the full range of $m$. It does not hold on every protein: on protein~G~B1 the in-class and random arms separate at $m=6$ and $m=10$ but not at $m=20$, where both sit within roughly $3\times$ chance (Figure~\ref{fig:contacts}, bottom row).

Figure~\ref{fig:mechanism}(c) asks whether the model simply finds relabeled sequences unfamiliar. We measure unfamiliarity as the drop in the model's log-likelihood, $\Delta_{\mathrm{LL}} = \mathrm{LL}(x) - \mathrm{LL}(\pi(x))$, in nats per residue. Over the $144$ random and in-class cells, $\Delta_{\mathrm{LL}}$ and overlap are correlated (Spearman $-0.700$; $-0.551$ with the cross-class arm included). Adding a regime indicator still raises $R^2$ from $0.491$ to $0.621$. In the random arm, $\Delta_{\mathrm{LL}}$ saturates by $m=6$ while overlap keeps falling by another order of magnitude, so unfamiliarity alone does not explain the collapse.

\subsection{RQ3: Does Scale Close The Gap?}
\label{sec:rq3}

\begin{figure}[t]
  \centering
  \includegraphics[width=0.8\textwidth]{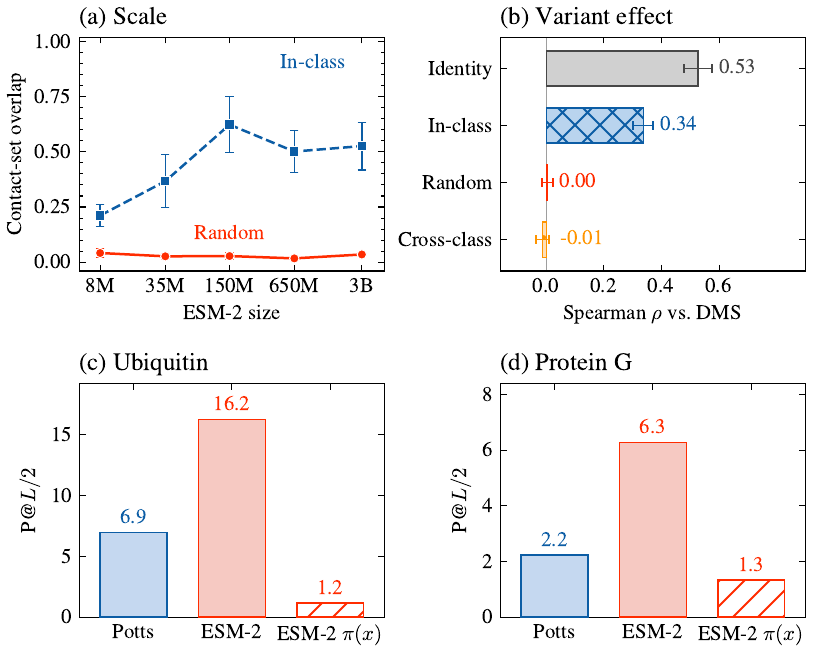}
  \caption{\textbf{The degradation does not shrink with scale.} \textbf{(a)} Overlap at nominal $m{=}20$ against parameter count. \textbf{(b)} shows Spearman $\rho$ against deep mutational scanning measurements on the $11$ ProteinGym assays where ESM-2 has signal. A PSSM on the same assays is unchanged in every cell. \textbf{(c)} and \textbf{(d)} show True long-range P@$L/2$ divided by contact density. On both ubiquitin and Protein G, ESM-2 650M degrades significantly under $\pi(x)$, falling below Potts.}
  \label{fig:cost}
\end{figure}

Scale does not close the gap. Across the five ESM-2 sizes, the random arm stays near zero, from $0.042$ at 8M to $0.035$ at 3B, while the in-class arm climbs, so the gap between them grows with scale (Figure~\ref{fig:cost}a). The pattern is consistent with larger models becoming more sensitive to biochemical similarity, while arbitrary relabeling remains destructive at every size.

The degradation also costs accuracy. Figure~\ref{fig:cost}(b) restricts both methods to the same candidate pairs so that their precisions are directly comparable. On ubiquitin, ESM-2 650M starts at $2.33\times$ the Potts precision ($0.737$ against $0.316$). Under relabeling it falls to $0.053$, while Potts is unchanged. On protein~G the same pattern is weaker. Unpermuted, ESM-2 reaches precision $0.763$ on full-length ubiquitin, so the loss is from an accurate model.

The failure is not specific to the categorical Jacobian. It replicates on variant-effect prediction, a second readout with external ground truth, over the $11$ of $15$ ProteinGym single-mutant assays where ESM-2 has signal (Figure~\ref{fig:cost}c; \citealp{notin2023proteingym}). Random relabeling at $m=20$ lowers Spearman $\rho$ by $0.52$ on average, in-class by $0.19$ and cross-class by $0.54$, while a PSSM baseline scored on the relabeled alignment is exactly unchanged in every cell (Appendix~\ref{app:readout2}). It also replicates under ESM-2's own supervised contact head on the same nine proteins and relabelings, where random relabeling drives overlap to $0.028$ at $m=20$ and in-class holds $0.793$ (Appendix~\ref{app:controls}).

\section{Discussion}
\label{sec:discussion}

\subsection{Alphabet anchoring vs. coevolutionary inference}

Our findings show that contact prediction in pLMs is driven by residue-identity-conditioned pattern matching rather than by the direct extraction of column statistics. We call this \textbf{alphabet anchoring}: the readout holds when residues are swapped within biochemical classes and collapses when swaps cross class boundaries. The term follows work in natural language processing on models that adopt surface heuristics in place of systematic statistical computation \citep{mccoy2019right}.
\newpage
We also rule out several competing accounts:
\begin{enumerate}[label=(\roman*)]
    \item \emph{Not pure column-statistic inference:} A model performing algebraic inversion of covariation would remain invariant across all three regimes (as proven for Potts models in Propositions~1 and~2).
    \item \emph{Not input difficulty (Hamming distance):} When edit distance is held constant by the matched-Hamming control on cytochrome~c, in-class permutations keep $54\times$ more contacts than either matched substitution, while random substitutions collapse to noise.
    \item \emph{Not an artifact of one architecture or readout:} The dissociation replicates across single-sequence models (ESM-2), alignment-based architectures (MSA Transformer), alternative tokenizers and vocabularies (AMPLIFY), and a supervised contact head (Appendix~\ref{app:controls}).
\end{enumerate}
Alphabet anchoring also explains why cross-class relabeling degrades contacts faster than random relabeling at $m=6$ and $m=10$ (Table~\ref{tab:main}). A random derangement keeps some residues inside their biochemical group by accident (Figure~\ref{fig:regimes}), whereas a cross-class mapping moves every one of them out.

\subsection{Out-of-distribution shift and scale}
\label{sec:ood}

A natural objection is that relabeling pushes sequences out of the model's training distribution, which would make any degradation unremarkable. We measure how unfamiliar the model finds a sequence with its log-likelihood shift $\Delta_{\mathrm{LL}}$, a zero-shot measure of sequence fitness \citep{meier2021language}. Unfamiliarity and overlap are correlated (Spearman $-0.70$), but unfamiliarity alone does not explain the collapse. In the random arm, $\Delta_{\mathrm{LL}}$ stops rising after $m{=}6$ while overlap keeps falling by another order of magnitude (Figure~\ref{fig:mechanism}c). In the in-class arm, $\Delta_{\mathrm{LL}}$ keeps rising while ESM-2 650M's overlap stays above $0.65$, so a sequence the model finds unfamiliar can still keep most of its contact map when biochemical classes are intact. After accounting for $\Delta_{\mathrm{LL}}$, regime still explains additional variance in overlap ($\Delta R^2 = 0.131$, partial Spearman $+0.506$). Throughout, the information a column-statistics estimator uses is unchanged, and Potts keeps its precision on the same inputs.

Scale does not remove the effect either. Increasing ESM-2's size $375\times$, from 8M to 3B parameters, improves tolerance to in-class swaps and gives no improvement under random relabeling (Figure~\ref{fig:cost}a). Language-model loss improves predictably with scale \citep{kaplan2020scaling}; for contact readouts, scale refines biochemical pattern recognition without making the readout invariant.

\subsection{Biochemical class as a categorical prior}
\label{sec:negative}

A graded substitution cost does not reliably predict degradation beyond the discrete class partition. Across full $m{=}20$ derangements, the correlation between BLOSUM62 substitution cost and overlap is $\rho = -0.43$ on ubiquitin and $-0.11$ on protein~G. pLMs therefore appear to use categorical groupings rather than a continuous distance over residues.

How much biochemical class protects the map depends on the protein. In-class relabeling at nominal $m{=}20$ keeps $0.675$ of the identity contacts on ubiquitin and $0.071$ on protein~G, where the in-class and random arms still separate at $m=6$ and $m=10$. Random relabeling collapses the map on every protein we test; what varies is how much of it biochemical class can save.

\section{Conclusion}

Contact maps derived from protein language models are widely interpreted as evidence that these models have internalized coevolution. In this work, we tested that interpretation directly, rather than through accuracy, using a property of any estimator that reads only column statistics: its contacts are invariant under a global relabeling of the amino-acid alphabet. We proved this for pseudolikelihood Potts models and as an exact identity for the inverse-covariance estimator. Every pLM we evaluated changes its contact predictions under this label-free, structure-free and alignment-free test, across three architectures, two tokenizer schemes and a $375\times$ range of scale, while the inverse-covariance estimator returns an identical score matrix on the same inputs. On ubiquitin, over a candidate set that allows direct comparison, ESM-2 goes from $2.33\times$ the Potts precision to $0.17\times$ under relabeling.

We call this phenomenon \textbf{alphabet anchoring}. While it does not undermine the utility of pLM contact maps, we show that it changes their mechanistic interpretation. A scaling curve of contact precision mixes coevolutionary inference with amino-acid pattern matching, so high precision cannot be cited as proof that a model computes contacts through coevolution. Where pLMs have replaced classical estimators, Figure~\ref{fig:cost}(b) is a case for keeping classical coevolutionary methods in the loop.

This work is a first step toward finding such pitfalls in protein language models. As pLMs are deployed across biological applications, the scientists and practitioners who rely on them should know that their contact maps depend on canonical residue identity. We hope this test inspires further diagnostics that check what a model computes in addition to how accurately it predicts.

\section*{Limitations}
\label{sec:limitations}

\paragraph{Scope:}
Our experiments are carried on a relatively small number of proteins, four Pfam families, three readouts (categorical-Jacobian contacts, ESM-2's supervised contact head, and ProteinGym variant effects), and three of the many available pLM families. The proteins are short and single-domain, the setting where classical coevolutionary estimators are often strongest and where the coevolution interpretation of pLMs is most often invoked. We did not test multi-domain proteins or shallow families. The in-class regime uses one grouping, derived from BLOSUM62; groupings from other substitution matrices, such as BLOSUM45 \citep{henikoff1992amino} or PAM \citep{dayhoff1978model}, were not tested.

\paragraph{The test is a necessary, but not sufficient evidence:}
Invariance under alphabet relabeling is a necessary condition for a contact readout to be a pure function of column statistics, but it is not sufficient to show that a model has learned coevolutionary inference. Our interventions isolate alphabet anchoring; they do not identify its cause. We also did not test retrained or fine-tuned models. A pLM trained on relabeled corpora might pass the test, but our claim concerns the public pre-trained checkpoints, which are the ones used in practice.

\section*{Reproducibility, Code, and Data Availability}

We release all of the code and data used in this paper at \url{https://github.com/aryankeluskar/icbinb-alphabet-anchoring}. No newly collected data was used in this research.

\section*{Disclosure of LLM Use}
In accordance with the NeurIPS policy on LLM usage, we disclose that Claude Code was used to assist with initial drafting and refining sentences in this paper. All content underwent multiple rounds of extensive revision, editing and verification by the authors. None of the citations or references were created by an LLM, instead all such content was manually checked and entered by the authors.

\label{endofbody}

\bibliographystyle{plainnat}
\bibliography{refs}

\newpage

\appendix

\section{Invariance proofs}
\label{app:proofs}
\label{app:preliminaries}

\paragraph{Shared contact pipeline.}
Potts models and pLMs both yield a coupling tensor of shape $L \times q \times L \times q$: Potts from its fitted couplings $J_{ij}$ or a regularized precision matrix, and a pLM from its categorical Jacobian $J[i,a,j,b]$, the change in the logit for residue $b$ at position $j$ when position $i$ is set to residue $a$~\citep{zhang2024protein}. The same pipeline turns either into an $L \times L$ contact score $S$: mean-centering over both amino-acid axes, symmetrization, the Frobenius norm of each $q \times q$ block, and average-product correction (APC)~\citep{dunn2008mutual}. For Potts, every input to this pipeline is a function of the alignment's column statistics, which is what makes the symmetry below available.

\paragraph{Notation.}
An alignment is $X \in \bar{\mathcal{A}}^{N \times L}$ over $\bar{\mathcal{A}} = \mathcal{A} \cup \{-\}$, where $\mathcal{A}$ is the $20$ standard amino acids and $-$ is the gap ($q = 21$ states; non-standard characters are mapped to the gap). A relabeling is a bijection $\pi: \mathcal{A} \to \mathcal{A}$, extended to $\bar{\mathcal{A}}$ by $\pi(-) = -$ and applied entrywise, $(\pi X)_{ni} = \pi(X_{ni})$. Let $P_\pi \in \{0,1\}^{q \times q}$ with $(P_\pi)_{a,\pi(a)} = 1$, so that $e_a^\top P_\pi = e_{\pi(a)}^\top$ and $P_\pi^\top P_\pi = I$.

\paragraph{Proposition 1 (pseudolikelihood Potts).}
\emph{Let $(h, J)$ maximize the regularized pseudolikelihood of $X$ under $H(\sigma) = -\sum_i h_i(\sigma_i) - \sum_{i<j} J_{ij}(\sigma_i, \sigma_j)$ with penalty $\lambda_J \sum_{i<j} \|J_{ij}\|_F^2 + \lambda_h \sum_i \|h_i\|_2^2$, $\lambda_h \geq 0$. Define $h^\pi_i = P_\pi^\top h_i$ and $J^\pi_{ij} = P_\pi^\top J_{ij} P_\pi$, that is, $h^\pi_i(\pi a) = h_i(a)$ and $J^\pi_{ij}(\pi a, \pi b) = J_{ij}(a, b)$. Then $(h^\pi, J^\pi)$ maximizes the regularized pseudolikelihood of $\pi X$, the attained objective is identical, and the contact scores are unchanged.}

\emph{Proof.}\quad By construction $H^\pi(\pi\sigma) = H(\sigma)$ for every $\sigma \in \bar{\mathcal{A}}^L$. Each conditional $p(\sigma_i \mid \sigma_{\setminus i})$ normalizes over $\sigma_i \in \bar{\mathcal{A}}$, which $\pi$ permutes, so $p^\pi(\pi\sigma_i \mid \pi\sigma_{\setminus i}) = p(\sigma_i \mid \sigma_{\setminus i})$, and the pseudolikelihood of $\pi X$ under $(h^\pi, J^\pi)$ equals that of $X$ under $(h, J)$ term by term. The penalty is unchanged because permutation matrices preserve Euclidean and Frobenius norms. The map $(h, J) \mapsto (h^\pi, J^\pi)$ is a bijection on parameter space (its inverse is the same map for $\pi^{-1}$) that preserves the objective, so it carries maximizers to maximizers. Finally, the zero-sum gauge and the scoring pipeline commute with $J_{ij} \mapsto P_\pi^\top J_{ij} P_\pi$ by the argument in the proof of Proposition~2, so the contact scores are identical. $\square$

This covers plmDCA, GREMLIN and CCMpred~\citep{ekeberg2013improved, seemayer2014ccmpred}, provided the penalty is the same for every amino acid.

\paragraph{Proposition 2 (inverse-covariance Potts).}
\label{sec:prop2}
\emph{Let $W = (C + \lambda I)^{-1}$ be the regularized precision matrix of the one-hot encoded alignment, and let $S$ be the contact score after mean-centering, symmetrization, per-block Frobenius norm, and APC. Then $S^\pi = S$ exactly.}

\emph{Proof.}\quad One-hot encode the alignment as $Y \in \mathbb{R}^{N \times qL}$, blocked by position. Relabeling permutes the $q$ coordinates of every block identically: $Y^\pi = YQ$ with $Q = I_L \otimes P_\pi$, which is orthogonal. The empirical covariance transforms by congruence, $C^\pi = Q^\top C Q$, so
\[
W^\pi = (Q^\top C Q + \lambda I)^{-1} = \big(Q^\top (C + \lambda I) Q\big)^{-1} = Q^\top (C + \lambda I)^{-1} Q = Q^\top W Q,
\]
using $Q^\top Q = I$. Blockwise, $W^\pi_{ij} = P_\pi^\top W_{ij} P_\pi$. Each scoring step commutes with this conjugation: mean-centering (row and column means are permuted, not changed), symmetrization ($(Q^\top W Q)^\top = Q^\top W^\top Q$), the per-block Frobenius norm ($\|P_\pi^\top W_{ij} P_\pi\|_F = \|W_{ij}\|_F$), and APC (a function of the already unchanged $L \times L$ matrix). Hence $S^\pi_{ij} = S_{ij}$. $\square$

\paragraph{Both conditions are necessary.}
The proof needs an isotropic ridge ($Q^\top \Lambda Q \neq \Lambda$ for a non-uniform $\Lambda$) and a gap that $\pi$ leaves fixed. We tested each in both directions on an alignment with $N{=}1500$, $L{=}40$ and random $m{=}20$. A per-amino-acid ridge spanning only $1.0$--$1.5\times\lambda$ gives max$|\Delta S| = 0.355$ ($11.8\%$ of the score scale; top-$L/2$ overlap $0.850$), against $4.9 \times 10^{-15}$ with an isotropic ridge. Letting the gap be relabeled (gap $\to$ K, with $10\%$ gaps and $1.5\%$ X) gives max$|\Delta S| = 0.182$, against $1.8 \times 10^{-15}$ when the gap is fixed.

\section{Pseudolikelihood invariance}
\label{app:c13}

The classical models a reader has in mind, plmDCA, GREMLIN and CCMpred, are pseudolikelihood fits with a real optimiser rather than the closed-form estimator of Proposition~2. We fitted one (asymmetric pseudolikelihood, per-column multinomial logistic regression, isotropic $L_2$, L-BFGS, zero-sum gauge, symmetrised, Frobenius per block, APC) on the same MSAs and targets.

On ubiquitin ($L{=}67$, depth $3000$, $\lambda{=}0.001$), the correctness gate (plm P@$L/2 \geq$ mean-field P@$L/2$) passes at $0.515$ against $0.333$. Under relabeling, a zero (equivariant) initialisation returns an identical top-$L/2$ contact set with identical precision, and a random initialisation returns $0.970$ overlap. Protein~G replicates: zero init gives $1.000$ overlap, random init $1.000$.

The raw score matrix diverges at $1.9 \times 10^{-3}$ relative, which is not machine precision. A maxiter ladder shows why: the two fits agree to $1.1 \times 10^{-15}$ at two iterations, $1.4 \times 10^{-14}$ at ten, $1.2 \times 10^{-9}$ at forty, and $1.9 \times 10^{-3}$ at $150$, twelve orders of magnitude of monotone growth, with the contact set exactly identical at every rung. The relabeled design matrix accumulates inner products in a different order, so gradients differ at $\sim 10^{-16}$ from the first step, and L-BFGS is not contractive. The theorem is about the computation, and a numerical solver inherits it only up to its own conditioning. Score-level invariance to machine precision belongs to the closed-form estimator alone; prediction-level invariance belongs to both.

\section{Experimental setup}
\label{app:details}

\paragraph{Proteins.}
Table~\ref{tab:proteins} lists the nine proteins. True-precision targets (Figure~\ref{fig:cost}b) are ubiquitin (1UBQ\_A, $111$ contacts at $|i-j| \geq 12$) and protein~G~B1 (1PGB\_A, $79$ contacts), the only two whose sequences are identical to their deposited chains; for example, our cytochrome~c is human while 1HRC is horse heart ($88.5\%$ identity).

\begin{table}[h]
  \centering
  \small
  \begin{tabular}{lcll}
    \toprule
    Protein & $L$ & Structure & Also used in \\
    \midrule
    PF00018 (BOI2\_YEAST) & $51$ & -- & S, M \\
    protein G B1 & $56$ & 1PGB & S, P, H, A \\
    PF00013 (B3RHY5\_TRIAD) & $67$ & -- & -- \\
    PF00076 (LA\_HUMAN) & $69$ & -- & M \\
    ubiquitin & $76$ & 1UBQ & S, P, H, A \\
    PF00027 (P73234\_SYNY3) & $89$ & -- & M \\
    cytochrome c & $104$ & 1HRC & S, H, A \\
    PF00072 (AGMR\_PSEAE) & $114$ & -- & M \\
    T4 lysozyme & $164$ & 2LZM & -- \\
    \bottomrule
  \end{tabular}
  \vspace{6pt}
  \caption{Proteins. All nine are used for the ESM-2 650M sweep (Table~\ref{tab:main}) and the contact-head replication (Appendix~\ref{app:controls}). Also used in: S, scale sweep; P, precision against structure; H, matched-Hamming control; A, AMPLIFY; M, MSA Transformer (full Pfam family alignment). Pfam alignments are from \citet{mistry2021pfam}, structures from \citet{berman2002protein}.}
  \label{tab:proteins}
\end{table}

\begin{figure}[t]
    \centering
    \includegraphics[width=\textwidth]{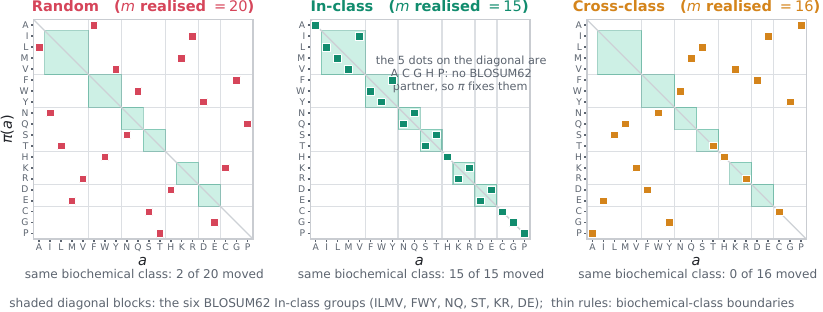}
    \caption{Each panel is the seed-$0$, $m{=}20$ permutation matrix, with a marker at $(a, \pi(a))$. Axes are ordered by biochemical class, so the six BLOSUM62 in-class groups form shaded blocks on the diagonal. Random preserves class only by accident ($2$ of $20$ residues here, $2$--$4$ across seeds). In-class stays inside the blocks; the five diagonal markers are A, C, G, H and P, which have no BLOSUM62 partner. In cross-class, none of the $16$ moves stays within a BLOSUM62 group. This is why the cross-class arm can fall faster than the random arm at the same nominal $m$.}
    \label{fig:regimes}
\end{figure}

\paragraph{Relabeling regimes.}
Permutations are generated by \texttt{src/permute.py} with NumPy seed $1000 \cdot \text{seed} + m$. At $m \in \{2, 6, 10\}$ every regime moves exactly $m$ residue types. At nominal $m{=}20$, in-class moves $15$ (A, C, G, H and P have no BLOSUM62 partner and stay fixed) and cross-class moves $14$ to $16$. Figure~\ref{fig:regimes} shows the seed-$0$ maps. Tables report nominal $m$; figures plot realized $m$.

\paragraph{Matched-Hamming control.}
\label{app:c6}
For each (protein, regime, $m$, seed) we build three sequences that differ from wild type at the same set of positions: the relabeled sequence, a uniform substitution (replacement drawn from the other $19$ residues), and a composition-matched substitution (replacement drawn from the wild type's own residue composition). The arms differ only in which residues they install; identical positions were asserted in every run. On cytochrome~c at in-class $m{=}20$ (Figure~\ref{fig:mechanism}a; $76$ of $104$ positions changed, $k{=}52$), the relabeled sequence keeps $37$, $37$ and $35$ of its $52$ contacts across seeds, the uniform arm $0$, $1$ and $1$, and the composition-matched arm $1$, $0$ and $1$.

\paragraph{Models and readouts.}
ESM-2 contacts use the categorical Jacobian with BOS and EOS replaced by the mask token, reduced as in Appendix~\ref{app:proofs}. The scale sweep uses ESM-2 8M, 35M, 150M, 650M and 3B on four proteins with two seeds ($n{=}8$ per cell). MSA Transformer uses four Pfam families with three seeds ($n{=}12$ per cell; $n{=}4$ for cross-class at $m{=}20$); inverse-covariance Potts on the same relabeled alignments returns overlap $1.000$ in all $136$ runs. AMPLIFY 120M (2024 checkpoint) runs in fp32 through the same post-processing and passed a validity gate before any relabeled run (unrelabeled P@$L/2 \geq 3\times$ chance; $4.4\times$ and $4.0\times$ on the two precision targets). Inverse-covariance Potts uses an isotropic ridge $\lambda = 4.5/\sqrt{N}$; for the precision comparison, alignment rows with more than $50\%$ gaps in the covered region are removed.

\paragraph{Log-likelihood shift.}
$\mathrm{LL}(s) = \frac{1}{L}\sum_i \log \mathrm{softmax}(\mathrm{logits}(s)_i)[s_i]$, with logits restricted to the $20$ standard residues and BOS/EOS masked, and $\Delta_{\mathrm{LL}} = \mathrm{LL}(x) - \mathrm{LL}(\pi(x))$. Its scale differs between models, so Figure~\ref{fig:mechanism}(c) overlays model sizes to compare the shape of each curve (where it saturates), not its level.

\paragraph{Metrics.}
Two residues are in contact when their C$\beta$ atoms (C$\alpha$ for glycine) are within $8$\,\AA. Candidate pairs have $|i - j| \geq 12$, below the CASP long-range cutoff of $24$, because the proteins are short ($L = 51$ to $164$); results at $|i - j| \geq 24$ are reported in Appendix~\ref{app:controls}. Overlap is $|A \cap B| / k$, where $A$ and $B$ are a model's top-$k$ candidate pairs before and after relabeling and $k = \lfloor L/2 \rfloor$. This is the overlap coefficient; it is not the Jaccard index $|A \cap B| / |A \cup B|$, which is lower whenever $0 < |A \cap B| < k$. Chance overlap is $k$ divided by the number of candidate pairs.

\paragraph{Variant effect.}
\label{app:readout2}
We use $15$ ProteinGym single-mutant assays~\citep{notin2023proteingym} with three seeds per cell. ESM-2 scores are masked marginals~\citep{meier2021language}: a mutation $a_i \to b$ is scored as $\log p(\pi(b)) - \log p(\pi(a))$ at masked position $i$ of the relabeled wild type, and the measured fitness values are unchanged. Unrelabeled Spearman $\rho$ reproduces the published ESM-2 650M values to four decimals on all $15$ assays. A PSSM built from the relabeled alignment and scored on the relabeled mutations is unchanged to floating-point zero in every cell. Eleven assays have unrelabeled $\rho \geq 0.20$; on these, random $m{=}20$ lowers $\rho$ by $0.52$ on average, in-class by $0.19$, and cross-class by $0.54$. The other four reproduce the published near-zero baseline.

\section{Example sequences.}
\label{app:verbatim}
The seed-$0$, $m{=}20$ maps applied to ubiquitin ($L{=}76$), with Hamming distance from wild type:

{\scriptsize
\begin{verbatim}
  wild type        MQIFVKTLTGKTITLEVEPSDTIENVKAKIQDKEGIPPDQQRLIFAGKQLEDGRTLSDYNIQKESTLHLVLRLRGG
  random    d=76   EWNARMPTPFMPNPTGRGQCYPNGSRMLMNWYMGFNQQYWWITNALFMWTGYFIPTCYVSNWMGCPTDTRTITIFF
  in-class  d=64   VNLWIRSMSGRSLSMDIDPTESLDQIRARLNERDGLPPENNKMLWAGRNMDEGKSMTEFQLNRDTSMHMIMKMKGG
\end{verbatim}
}

The in-class sequence occurs nowhere in nature and differs from wild type at $64$ of $76$ positions, yet ESM-2 keeps most of its contact map on it. Over all $2080$ candidate pairs of full-length ubiquitin ($k{=}38$, chance overlap $0.018$), random relabeling leaves ESM-2 650M with $1$, $0$ and $0$ of its $38$ contacts across seeds, and the rank correlation between the unrelabeled and relabeled score matrices is only $0.21$, $0.20$ and $0.13$, so the collapse is not limited to the top of the ranking.

\section{Additional results}
\label{app:controls}
\label{app:contacts}

\begin{figure}[h]
    \centering
    \includegraphics[width=\textwidth]{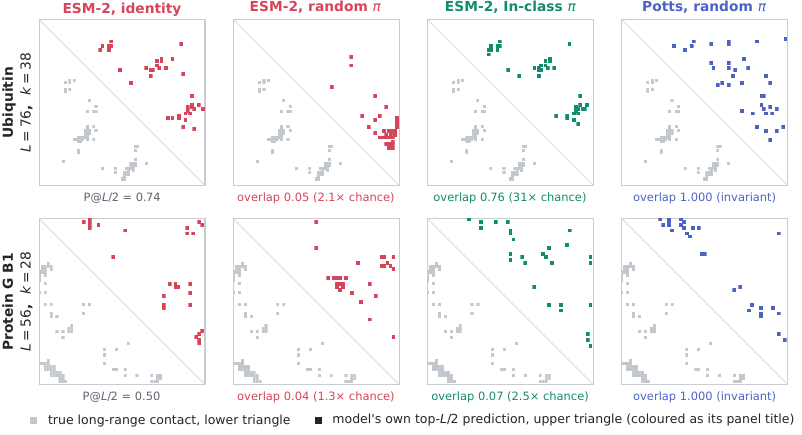}
    \caption{Contact maps for both precision targets on the restricted candidate set of Figure~\ref{fig:cost}(b). Lower triangle: contacts of the deposited chain. Upper triangle: the model's top-$L/2$ prediction. Overlaps are against the same model's unrelabeled prediction. Top row, ubiquitin: random relabeling destroys the map ($0.05$, $2.1\times$ chance) while in-class relabeling preserves most of it ($0.76$, $31\times$ chance). Bottom row, protein~G~B1: the in-class arm also collapses ($0.07$, $2.5\times$ chance), the counter-case discussed in \S\ref{sec:rq2} and \S\ref{sec:negative}. Potts returns its unrelabeled map exactly in both rows (overlap $1.000$, max$|\Delta S| \leq 2.3 \times 10^{-14}$).}
    \label{fig:contacts}
\end{figure}
\paragraph{Contact maps for both precision targets.}
Figure~\ref{fig:contacts} shows the maps behind Figure~\ref{fig:cost}(b), including protein~G~B1, where in-class relabeling does not protect the map.

\vspace{6pt}

\begin{figure}[h]
    \centering
    \includegraphics[width=\textwidth]{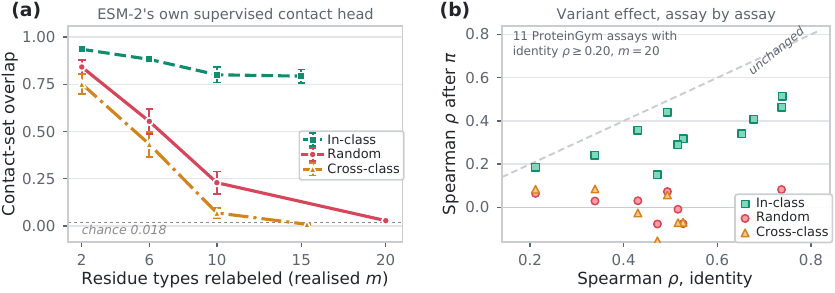}
    \caption{The failure is not specific to the categorical Jacobian or to one assay. \textbf{(a)} ESM-2 650M's supervised contact head; nine proteins, three seeds, mean $\pm$ SEM, plotted at realized $m$. \textbf{(b)} Variant-effect Spearman $\rho$ before and after relabeling at $m{=}20$, one marker per ProteinGym assay per regime; the dashed line is no change.}
    \label{fig:controls}
\end{figure}
\newpage
\paragraph{ESM-2's supervised contact head.}
The main results read contacts from the categorical Jacobian, so one objection is that the Jacobian, not the model, is what fails. We repeated the sweep with ESM-2 650M's own supervised contact head on the same nine proteins, regimes and seeds (Figure~\ref{fig:controls}a). It reproduces the dissociation: random relabeling falls from $0.842$ at $m{=}2$ to $0.028$ at $m{=}20$ ($1.5\times$ chance), cross-class to $0.009$, while in-class holds $0.793$. At $|i - j| \geq 24$ the result is unchanged ($0.025$ random, $0.774$ in-class at $m{=}20$). A head trained with structural supervision therefore inherits the same alphabet anchoring.

\end{document}